\documentclass{PoS}

\usepackage{amsmath}

\title{Pion electromagnetic form factor from full lattice QCD}

\ShortTitle{Pion electromagnetic form factor from full lattice QCD}

\author{HPQCD Collaboration}

\author{\speaker{Jonna Koponen}
        \\
        SUPA, School of Physics and Astronomy, University of Glasgow, Glasgow, G12 8QQ, UK\\
        E-mail: \email{jonna.koponen@glasgow.ac.uk}}

\author{Francis Bursa\\
        SUPA, School of Physics and Astronomy, University of Glasgow, Glasgow, G12 8QQ, UK}

\author{Christine Davies\\
        SUPA, School of Physics and Astronomy, University of Glasgow, Glasgow, G12 8QQ, UK}

\author{Gordon Donald\\
        SUPA, School of Physics and Astronomy, University of Glasgow, Glasgow, G12 8QQ, UK\\
        School of Mathematics, Trinity College, Dublin 2, Ireland}

\author{Rachel Dowdall\\
        DAMTP, University of Cambridge, Wilberforce Road, Cambridge CB3 0WA, UK}

\abstract{We present preliminary results from the first calculation of the pion
electromagnetic form factor at physical light quark masses. This form factor
parameterises the  deviations from the behaviour of a point-like particle when
a photon hits the pion. These deviations result from the internal structure of
the pion and can thus be calculated in QCD. We use three sets (different lattice
spacings) of $n_f=2+1+1$ lattice configurations generated by the MILC collaboration.
The Highly Improved Staggered Quark formalism (HISQ) is used for all of the
sea and valence quarks. Using lattice configurations with $u$/$d$ quark masses
very close to the physical value is an advantage, as we avoid the chiral
extrapolation. We study the shape of the vector ($f_+$) form factor in the
$q^2$ range from 0 to $-0.12~\textrm{GeV}^2$ and extract the mean square radius,
$\langle r^2_v\rangle$. The shape of the vector form factor and the resulting
radius is compared with experiment.}

\FullConference{31st International Symposium on Lattice Field Theory LATTICE 2013\\
		 July 29 -- August 3, 2013\\
		 Mainz, Germany}

\begin{document}

\section{Motivation}

\begin{figure}
\centering
\includegraphics[width=0.7\textwidth]{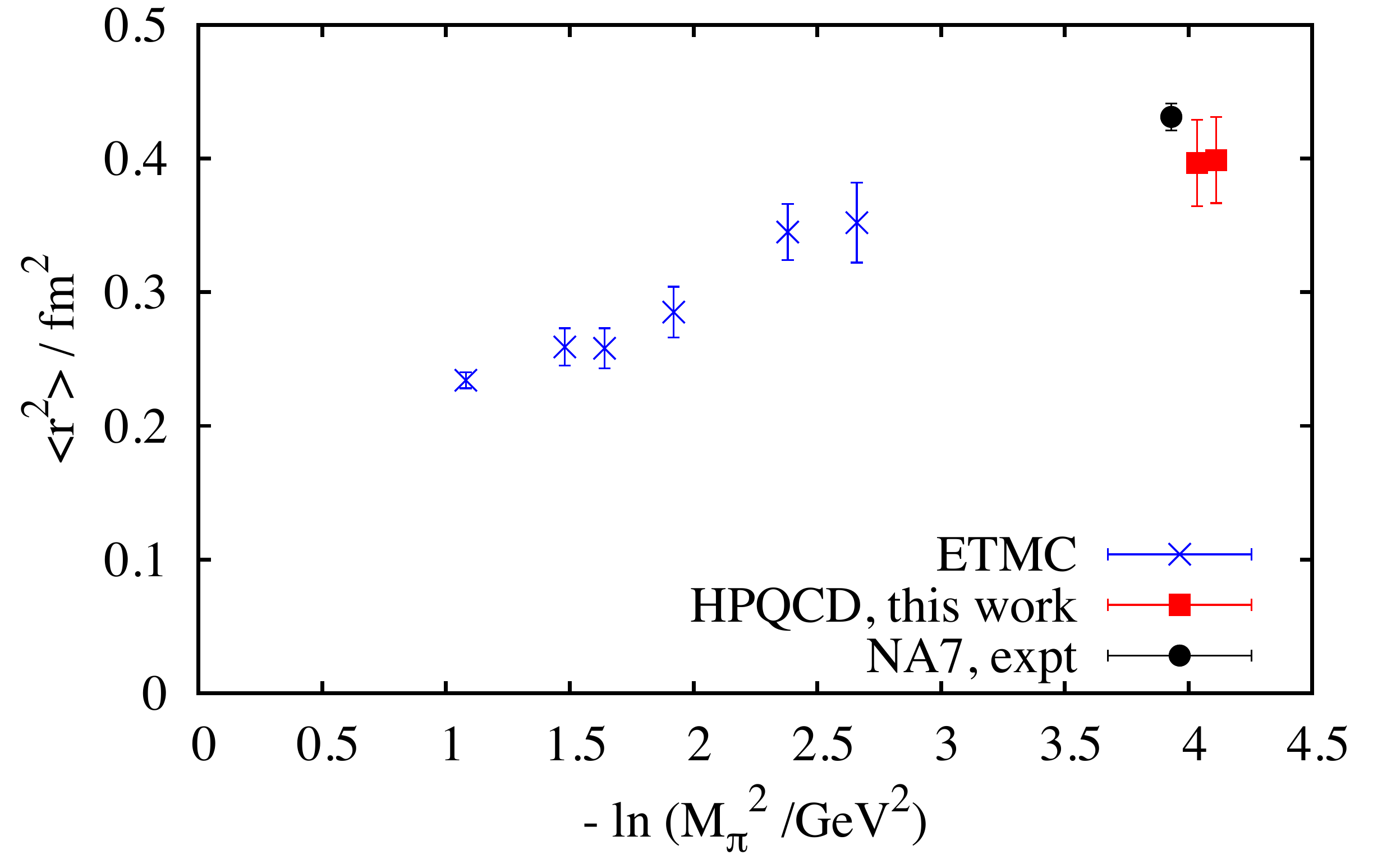}
\caption{Dependence of the charge mean square radius on the pion mass.
Note that we plot the radius as a function of $\ln (m_\pi^2)$, not $m_\pi^2$.
The experimental result is from~\cite{NA7} and the ETMC lattice results are
from~\cite{ETMC}.}
\label{fig:r2vsmass}
\end{figure}

The electromagnetic form factor of the charged $\pi$ meson parameterises the
deviations from the behaviour of a point-like particle when struck by a photon.
These deviations arise from the internal structure of the pion: constituent
quarks and their strong interaction. The form factor can be calculated in
Lattice QCD, but it is desirable to work at the physical pion mass to avoid chiral
extrapolation. Plots of lattice determinations of the pion charge radius as a
function of pion mass -- like Fig.~\ref{fig:r2vsmass}, or Fig.~11 in~\cite{ETMC}
or Fig.~4 in the very recent review paper~\cite{review} -- show very clearly
that the extrapolation to physical pion mass plays a key role, if the pion
masses one works at are much heavier than the physical mass. On the experimental
side, the vector form factor has been measured by NA7 collaboration \cite{NA7}
in a $\pi$ -- $e$ scattering experiment, which allows a comparison between
theory and experiment.

\section{Lattice configurations}

We use the $n_f=2+1+1$ HISQ (Highly Improved Staggered Quark) physical pion
mass lattice configurations provided by MILC Collaboration~\cite{MILCensembles}.
Three ensembles (different lattice spacings) are used in this study -- see details in
Table~\ref{table:configs}. We use the HISQ action for valence quarks as well,
and using the same light quark mass as in the sea. We have $Lm_\pi \approx 4$ for
the coarse ($a=0.12~\textrm{fm})$ and fine ($a=0.088~\textrm{fm}$) lattices,
so finite volume effects are expected to be very small. We have good
statistics with 1000 configurations (for the very coarse and coarse ensembles)
and four time sources per configuration.

\begin{table}
\centering
\begin{tabular}{|c|c|c|c|c|c|c|c|c|}
\hline
\hline
Set & $a$/fm & $am_l$ & $am_s$ & $am_c$ & $m_\pi$/MeV & $(L/a)^3 \times L_t/a$ & $N_\textrm{conf}$ & $T/a$\\
\hline
1 & 0.15  & 0.00235 & 0.0647 & 0.831 & 133 & $32^3 \times 48$ & 1000 & 9, 12, 15\\
2 & 0.12  & 0.00184 & 0.0507 & 0.628 & 133 & $48^3 \times 64$ & 1000 & 12, 15, 18\\
3 & 0.088 & 0.00120 & 0.0363 & 0.432 & 128 & $64^3 \times 96$ & 223  & 16, 21, 26\\
\hline
\hline
\end{tabular}
\caption{Details of the MILC 2+1+1 flavor lattice configurations used in this study.
Set 1 is very coarse, set 2 is coarse and set 3 is fine ensemble. The second column
is the lattice spacing (\cite{lattspacing}, using $w_0$ to determine
  the scale) and columns 3-6 list the sea quark masses and the pion mass.
Columns 7 and 8 give the size of the lattice and the number of configurations used.
Four time sources are used per configuration to get good statistics. $T$ in the 9th
column is the separation between the initial and final state
mesons. We use multiple values of $T$ to improve extraction of the
ground state matrix elements.}
\label{table:configs}
\end{table}

\section{Form Factors}

\begin{figure}
\centering
\includegraphics[width=0.80\textwidth]{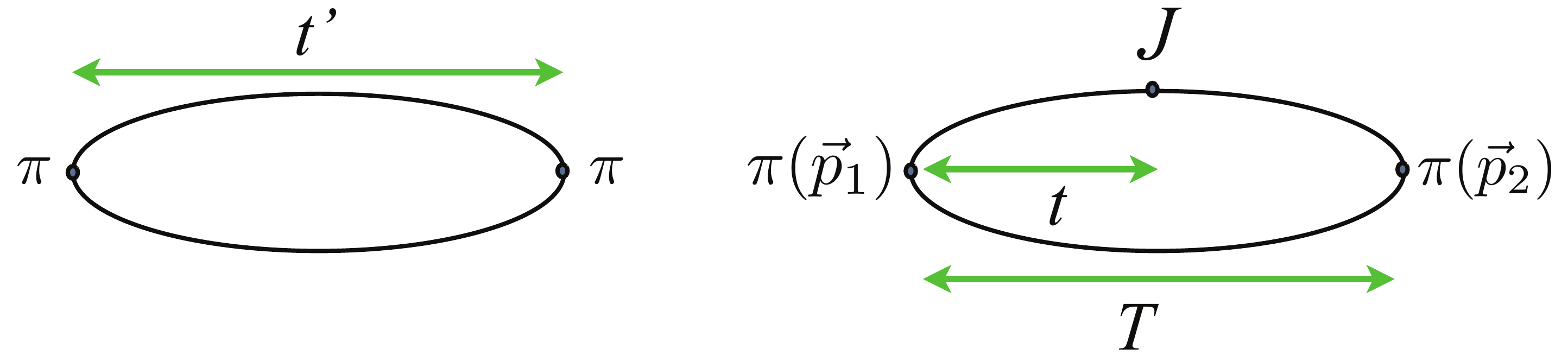}
\caption{2pt and 3pt correlation functions. $J$ denotes the current, $p_1$ is the
momentum of the initial pion and $p_2$ is the momentum of the pion in the final state.
Three different values of $T$, the separation between the initial and final state
mesons, were used in this study.}
\label{fig:diagrams}
\end{figure}

On the lattice, form factors are extracted from 3-point correlators -- see
Fig.~\ref{fig:diagrams}. We have two pseudoscalar mesons, the initial and final state 
pions, time $T$ apart, and we use a 1-link spatial vector current (in
the staggered formulation we need a 1-link
operator to make a taste singlet, as both pions are Goldstone mesons). A phase at the
boundary (twisted boundary condition,~\cite{twist}) is used to give the quarks momentum:
a twist
\begin{equation}
\Phi(x+\hat{e}_jL)=\textrm{e}^{i2\pi\theta_j}\Phi(x)
\end{equation}
is equivalent to the quark having a momentum
\begin{equation}
p_j=\frac{2\pi\theta_j}{L}.
\end{equation}
$\theta$ can be tuned to get the desired $q^2$, the four-momentum transfer defined
as
\begin{equation}
q^2=(E(\vec{p}_2)-E(\vec{p}_1))^2-(\vec{p}_2-\vec{p}_1)\cdot (\vec{p}_2-\vec{p}_1).
\end{equation}
We calculate the form factor $f_+(q^2)$ in the space-like (negative) region of $q^2$
near zero (this is the range where experimental data is available). In addition to
the 3-point correlators we also need the amplitudes (and energies) from the meson
2-point correlators.

\section{Fitting the correlators}

We fit the 2-point and 3-point correlators simultaneously. We use multi-exponential
fits to reduce systematic errors from the excited states, varying the number of 
exponentials. We take a fit that gives a good $\chi^2$ and check that increasing the
number of exponentials does not change the result. The preliminary results presented
in this paper are from a 5 exponential fit. Bayesian priors are used to constrain
the fit parameters. We fit all $q^2$ values simultaneously to take into account the
correlations. The fit functions for the 2-point correlators are of the form
\begin{equation}
C_{2pt}(p,t)=\sum_i b_i^2(\vec{p})\textrm{fn}(E_i(\vec{p}),t)
+\sum_i {b'_i}^2(\vec{p})\textrm{fo}(E'_i(\vec{p}),t)
\end{equation}
with
\begin{align}
\textrm{fn}(E,t)&=\textrm{e}^{-Et}+\textrm{e}^{-E(L_t-t)},\nonumber \\
\textrm{fo}(E,t)&=(-1)^{t/a}\bigl[\textrm{e}^{-Et}+\textrm{e}^{-E(L_t-t)}\bigr].
\end{align}
Here ``fn'' are the ``normal'' states and ``fo'' are the opposite parity states
that appear due to staggered quark formulation. The 3-point correlators are
fitted with
\begin{align}
C_{3pt}(\vec{p}_1,\vec{p}_2,t,T)=&\sum_{i,j} b_i(\vec{p}_1)\textrm{fn}(E_i(\vec{p}_1),t)
J^{nn}_{i,j}(\vec{p}_1,\vec{p}_2)b_j(\vec{p}_2)\textrm{fn}(E_j(\vec{p}_2),T-t)\\
-&\sum_{i,j} b_i(\vec{p}_1)\textrm{fn}(E_i(\vec{p}_1),t)
J^{no}_{i,j}(\vec{p}_1,\vec{p}_2)b'_j(\vec{p}_2)\textrm{fo}(E'_j(\vec{p}_2),T-t)
+ (n \leftrightarrow o).
\end{align}
Note that the amplitudes $b_i$ and the energies $E_i$, $E'_i$ are the same as in the
2-point correlators.

\section{Vector form factor}

\begin{figure}
\centering
\includegraphics[width=0.85\textwidth]{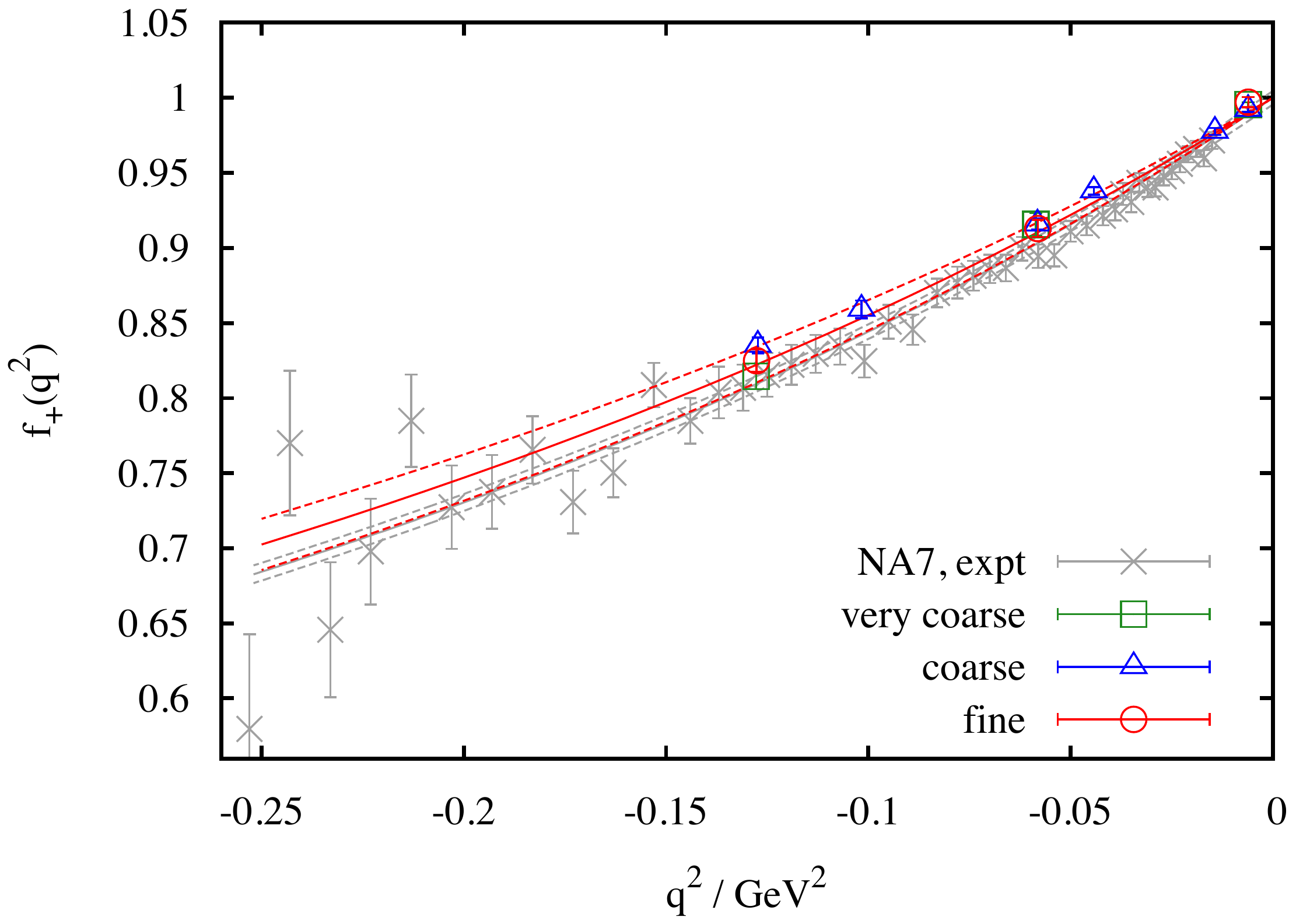}
\caption{Shape of the pion electromagnetic form factor. Experimental results
by NA7 Collaboration are from~\cite{NA7}, lattice results by HPQCD from this
work. The lines with error bands are from fits to experimental data (grey colour,
fit from~\cite{NA7}) and to our lattice results (red colour, this work).}
\label{fig:vectff}
\end{figure}

The matrix element relevant for the pion form factor is
\begin{equation}
\langle \pi (\vec{p}_1)|J|\pi (\vec{p}_2)\rangle=
Z\sqrt{4E_0(\vec{p}_1)E_0(\vec{p}_2)}J_{0,0}(\vec{p}_1,\vec{p}_2),
\end{equation}
where $J_{0,0}$ is the ground state amplitude of the 3-point correlator. The
form factor $f_+$ is related to the matrix element as
\begin{equation}
\langle \pi (\vec{p}_1)|V_i|\pi (\vec{p}_2)\rangle=
f_+(q^2)(\vec{p}_1+\vec{p}_2)_i.
\end{equation}
A renormalisation constant $Z$ is needed for the vector current: we normalise
the current by demanding that $f_+(0)=1$. Our results for the form factor
as a function of $q^2$ for different lattice ensembles are shown in
Fig.~\ref{fig:vectff} along with the experimental data points.

\section{Continuum extrapolation}

\begin{figure}
\centering
\includegraphics[width=0.95\textwidth]{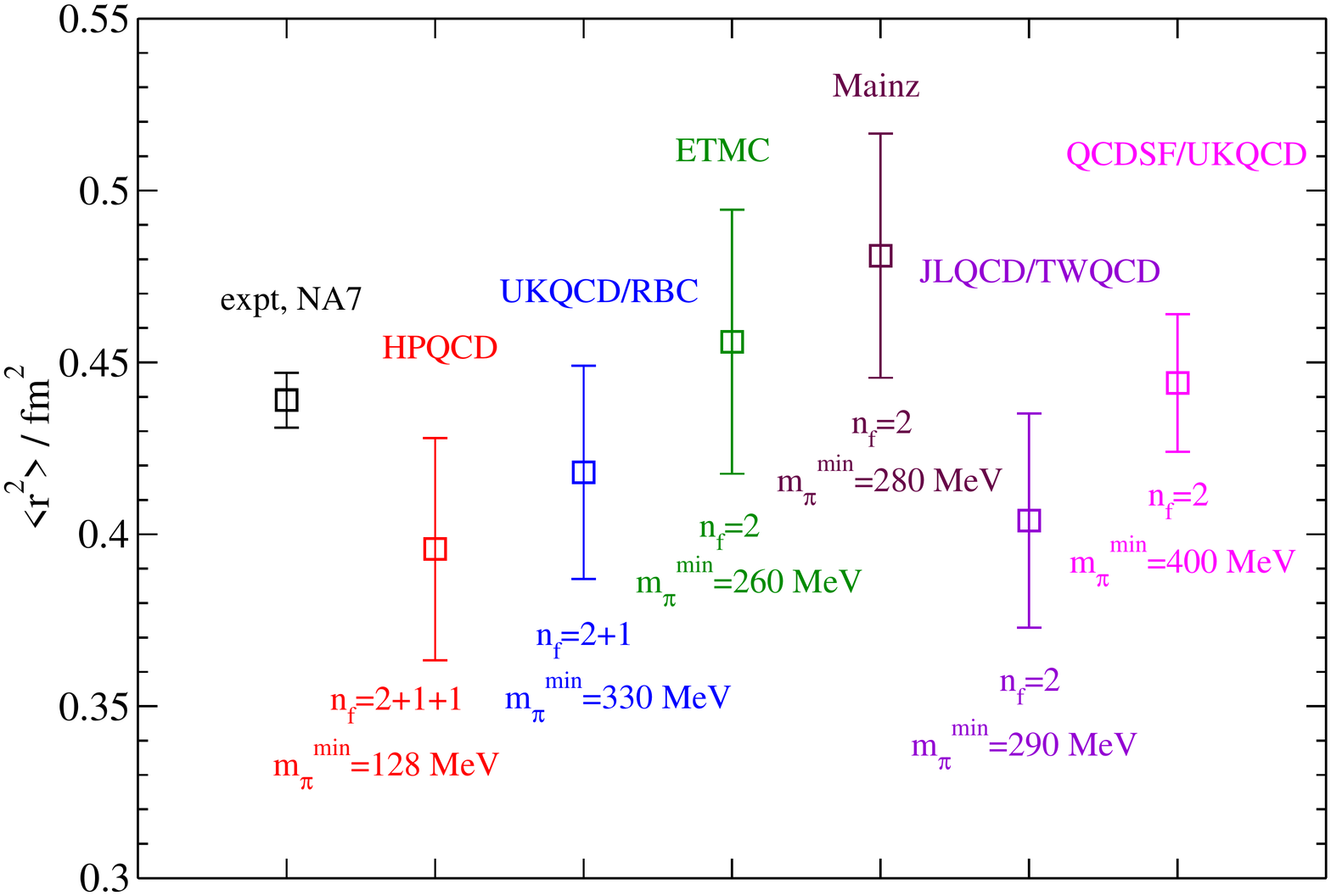}
\caption{Vector mean square radius. The experimental result is from~\cite{NA7},
and the HPQCD result is from this work. Other lattice results are from
\cite{UKQCD/RBC, ETMC, Mainz, JLQCD/TWQCD, QCDSF/UKQCD} (from left to right).
$n_f$ is the number of flavors and $m_\pi^{\textrm{min}}$ is the smallest pion mass
used in that calculation. The results shown here are each group's final result
after continuum and chiral extrapolation.}
\label{fig:rv2}
\end{figure}

We do a simultaneous extrapolation to zero lattice spacing and physical pion mass
by fitting the results from the very coarse, coarse and fine lattice to the pole
form 
\begin{equation}
f(q^2)=\frac{1}{1-q^2\langle r^2\rangle /6}
\end{equation}
allowing for $a^2$ and $m_\pi$ dependence:
\begin{equation}
\langle r^2\rangle =A(1+Ba^2+Ca^4)+c_J\ln (m_\pi^2/\mu^2),
\end{equation}
where $A, B, C$ are fit parameters and $c_J=1/(8\pi^2F^2)$ is a fixed
constant from NLO chiral perturbation theory (here $F$ is the pion
decay constant in the chiral limit). The form factor extrapolated to real
world is then the $a=0$, physical $m_\pi$ part of the fit function, also 
shown in Fig.~\ref{fig:vectff} with the error bands. The chiral log gives 
only a very small correction, as the pion masses are very
close to the physical value. The slope at $q^2=0$ gives the mean
square value of the charge radius:
\begin{equation}
\langle r^2_v\rangle =6\frac{\textrm{d}f_+(q^2)}{\textrm{d}q^2}|_{q^2=0}.
\end{equation}
Our preliminary result is $\langle r^2_v\rangle =0.40(3)~\textrm{fm}^2$. Comparison
to other lattice calculations and experiment in Fig.~\ref{fig:rv2} shows
good agreement.

\begin{figure}
\centering
\includegraphics[width=0.72\textwidth]{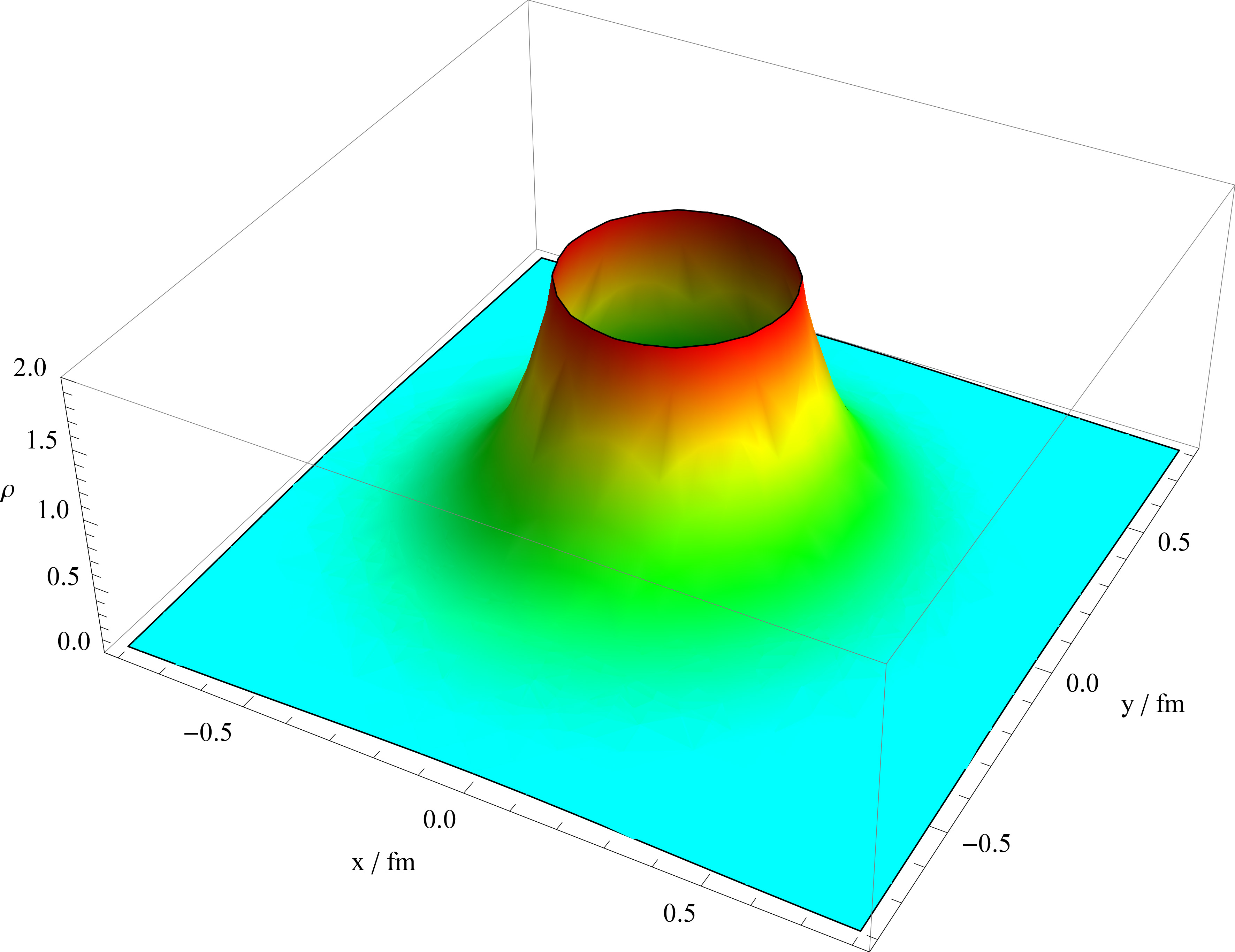}
\caption{Non-relativistic charge density calculated using the pole form and 
$\langle r^2_v\rangle$ from our fit. We omit $z$ direction
here for clarity and plot the charge density against $x$ and $y$.}
\label{fig:density}
\end{figure}

The form factor $f_+(q^2)$ can be viewed as a Fourier transform of the electric
charge distribution. Hence the charge density can be calculated from the vector
form factor once its functional form is known. In the non-relativistic limit
the charge density is
\begin{equation}
\rho(R) = \frac{3}{2\pi R\langle r^2_v\rangle}
\exp \Bigg(-\frac{R}{\sqrt{\langle r^2_v\rangle /6}}\Bigg)
\end{equation}
Using our result $\langle r^2_v\rangle =0.40~\textrm{fm}^2$ gives the charge
density plotted in Fig.~\ref{fig:density}.

\section{Summary}

We have presented here preliminary results from a full Lattice QCD calculation
of the pion vector electromagnetic form factor at physical pion mass. The use of
twisted boundary conditions allows us to calculate the form factor in the $q^2$
range from 0 to $-0.12 \textrm{ GeV}^2$ where experimental data is available. We
also determine the charge radius: our preliminary result is
$\langle r^2_v\rangle =0.40(3)~\textrm{fm}^2$. Comparison with experiment shows
very good agreement. 

\section*{Acknowledgements}

We are grateful to MILC for the use of their gauge configurations and
the MILC code. We used the Darwin Supercomputer in Cambridge as
part of the DiRAC facility jointly funded by
STFC, BIS and the Universities of Cambridge and Glasgow.

\end{document}